\documentclass[12pt]{article}

\usepackage{a4}
\usepackage{a4wide}

\usepackage{amsmath}
\usepackage{amscd}
\usepackage{amsfonts}
\usepackage{amssymb}
\usepackage{mathrsfs}
\usepackage{fancybox} 
\usepackage{enumerate}
\usepackage{cite}
\usepackage{bm}
\usepackage{amscd}
\usepackage{graphicx}
\usepackage[dvips]{color}
\usepackage{epsfig}
\usepackage{subfigure}

\makeatletter

\@addtoreset{equation}{section}
\makeatother
\if0
\usepackage{color}
\definecolor{blue1}{rgb}{0.5,0.15,0.10}
\definecolor{red1}{rgb}{0.8,0.1,0.10}
\usepackage[
debug,
dvipdfm,
colorlinks=true,
urlcolor=red1,
anchorcolor=red1,
citecolor=red1,
filecolor=red1,
linkcolor=red1,
menucolor=red1,
pagecolor=red1,
linktocpage=true,
pageanchor=false,
]{hyperref}    
\fi
\usepackage{amsthm} % This is for \newtheorem*
\def\half{{1\over2}}

\def\const{\mathrm{const}}

\def\arcsinh{\mathrm{arcsinh}}

\newcommand{\wt}{\widetilde}

\def\={\stackrel{\bullet}{=}}

\def\({\left(}
\def\){\right)}
\def\[{\left[}
\def\]{\right]}

\def\mbf{\mathbf}

\def \be {\begin{equation}}
\def \ee {\end{equation}}
\def \beqa {\begin{eqnarray}}
\def \eeqa {\end{eqnarray}}
\def \beal#1 {\begin{align}#1\end{align}}
\def \bes#1 {\begin{equation}\begin{split}#1\end{split}\end{equation}}
\def \nn {\notag\\}

\makeatletter

\@addtoreset{equation}{section}
\makeatother

\begin{document}

\begin{titlepage}
\title{
\vspace{-2cm}
\begin{flushright}
\normalsize{ 
YITP-21-142 \\ 
}
\end{flushright}
       \vspace{1.5cm}
Gravitational collapse of spherical shells of fluid in the isotropic homogeneous universe 
       \vspace{1.cm}
}
\author{
Shuichi Yokoyama\thanks{shuichi.yokoyama[at]yukawa.kyoto-u.ac.jp},\; 
\\[25pt] 
${}^{*}$ {\normalsize\it Center for Gravitational Physics,} \\
{\normalsize\it Yukawa Institute for Theoretical Physics, Kyoto University,}\\
{\normalsize\it Kitashirakawa Oiwake-cho, Sakyo-Ku, Kyoto 606-8502, Japan}
}

\date{}

\maketitle

\thispagestyle{empty}

% \vspace{.2cm}

\begin{abstract}
\vspace{0.3cm}
\normalsize

We investigate gravitational collapse of thick shell of fluid in the isotropic homogeneous universe without radiation described by the Einstein gravity with cosmological constant. 
We construct analytic solutions of this kind interpolating the Friedmann-Lemaitre-Robertson-Walker (FLRW) metric and the de-Sitter Schwarzschild black hole one by the Gullstrand-Painlev\'e one.
After determining the scale factor for a perfect fluid coexisting with cosmological constant, we determine the orbit of the collapsing shells of fluid from the continuity condition for density and the perpendicular component of pressure at interface. However the continuity condition cannot fix the orbit when there does not exist a perfect fluid inside the shells. In this case we determine the orbit from an equation of state for the fluid consisting a shell near the surface. 
Finally we confirm that the total energy defined in arXiv:2005.13233 is independent of the given time evolution in this system. 
\end{abstract}
\end{titlepage}
%\tableofcontents

\section{Introduction}
\label{Intro} 

The observation of a supermassive black hole and its event horizon \cite{2010RvMP...82.3121G,EventHorizonTelescope:2019dse} has made its existence manifest and endorsed that laws of gravitational physics are governed by General Relativity. 
A typical formation of a black hole is due to gravitational collapse, by which a compact object becomes too heavy to support itself by the internal repulsive force by the matter constituents.  
Therefore the investigation of gravitational collapse is highly motivated and it has been one of the traditional subjects in gravitational physics from an early period \cite{1939PhRv...56..455O,Datt1938,Misner:1964je,Penrose:1964wq,1965gtgc.book.....H,1966ApJ...144..943V,1966ApJ...143..682M,10.1143/PTP.38.92}. 
(See \cite{Joshi:2012mk} for a review and references therein.)

Another telescopic observation of high redshift supernova \cite{Riess:1998cb,Astier:2005qq} and cosmic microwave background \cite{Penzias:1965wn,Aghanim:2018eyx} indicates the existence of nonzero cosmological constant, whose value may be much smaller than the scale expected from particle physics \cite{Weinberg:1988cp}. 
By taking into account this result, it is motivated to investigate gravitational collapse in the universe with cosmological constant.   
There have been several works on gravitational collapse in the Einstein gravity with cosmological constant, for instance, \cite{1991GReGr..23..471G,Cissoko:1998mx,Yamanaka:1992zd,Markovic:1999di,Lake:2000rm,Madhav:2005kg,Ghosh:2006ab,Sharif:2006bp}, which were all studied by the so-called junction method \cite{MSM1927,zbMATH03073849,1955trdl.book.....L,10.2307/100525}.  
The junction method provides a general way to construct a metric to describe gravitational collapse by preparing two coordinate patches for the spacetime and gluing them smoothly at their overlap, though phenomena described by the glued metric may not be lucid as a whole.  

On the other hand, there is another approach to gravitational collapse by preparing for one good coordinate system to cover the entire space-time in which gravitational collapse happens \cite{Misner:1974qy,Landau:1982dva,Adler:2005vn}. 
For convenience we refer to this approach as the interpolating method. 
This method may be more refined than the junction method with regard that one needs to find out a good coordinate system which will smoothly interpolate the two coordinate systems glued by the junction method.  
In order to take the latter approach to studying gravitational collapse in our universe, one needs to prepare for a good coordinate system which interpolates the Friedmann-Lemaitre-Robertson-Walker (FLRW) metric and the de-Sitter Schwarzschild black hole metric. It was pointed out in \cite{Adler:2005vn} that such a coordinate system is given by the Gullstrand-Painlev\'e one \cite{gullstrand1922allgemeine,1921CR....173..677P}.
The Gullstrand-Painlev\'e coordinate system describes space-time like flows of a river flowing through a flat background \cite{Hamilton:2004au}, and it is useful to describe gravitational collapse of our interest. (See also \cite{Kanai:2010ae}.) 

\begin{figure}[th]
  \begin{center}
  \subfigure[Emergent black hole]{
  \includegraphics[scale=.45]{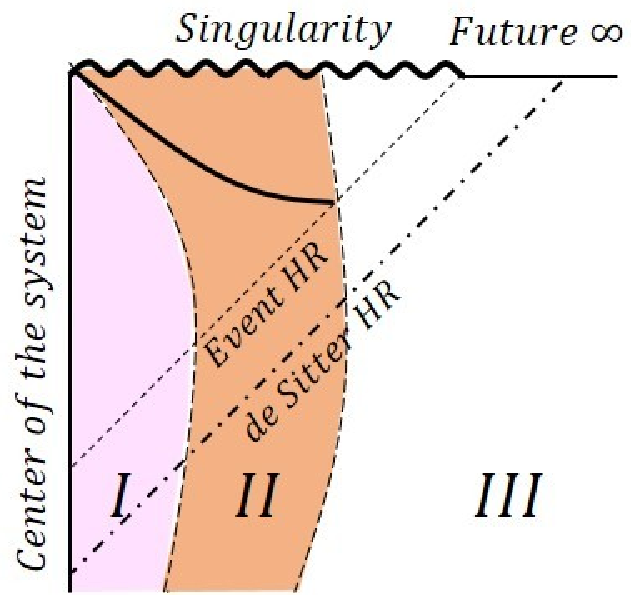}
  }
  \qquad
  \subfigure[Emergent local naked singularity 1]{
  \includegraphics[scale=.45]{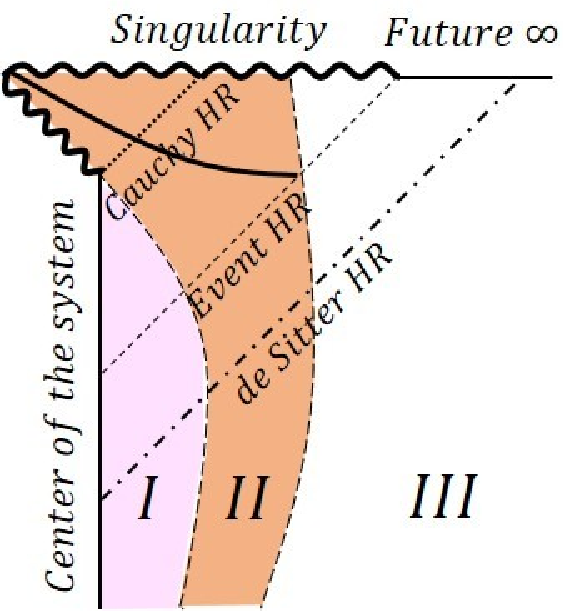}
  }
  \qquad
  \subfigure[Emergent local naked singularity 2]{
  \includegraphics[scale=.45]{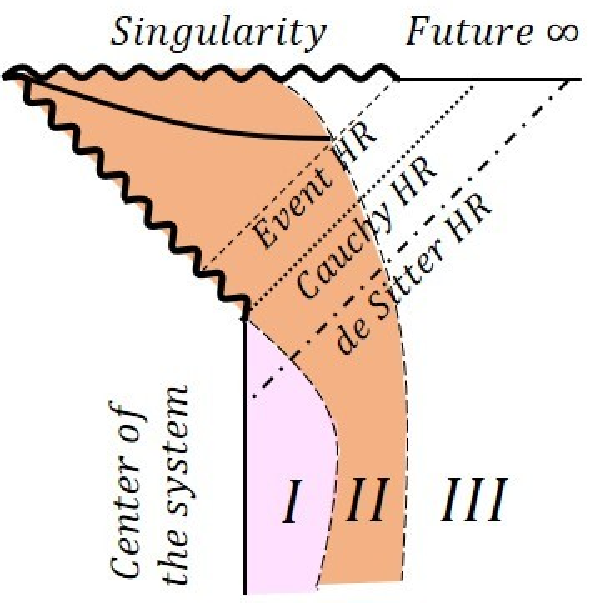}
  }
  \qquad
  \subfigure[Emergent global naked singularity]{
  \includegraphics[scale=.45]{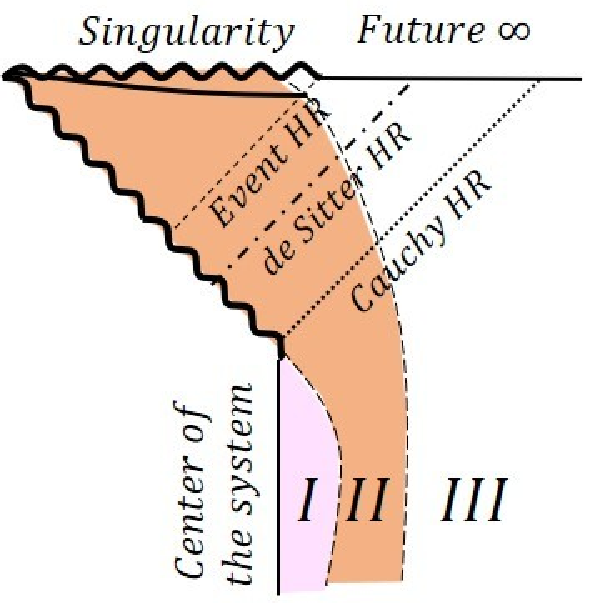}
  }
  \end{center}\vspace{-0.5cm}
  \caption{The Penrose-Carter diagrams to describe possible local causal structures of the gravitational collapse of thick shell of fluid are shown. ``HR'', ``Future $\infty$'' mean ``horizon'', ``Future infinity'', respectively. The bold line emanating from the singularity to the event horizon is the infinite redshift surface or the trapping horizon. The space-time is separated by the thick shell of fluid, whose region is denoted by II and covered by some color. The region inside, which is colored by another color and labeled by I, is filled with a perfect fluid, while in the region outside denoted by III there exists only cosmological constant or dark energy.  
   }
  \label{PC}
\end{figure}

The aim of this paper is to construct an analytic solution of a catastrophic gravitational collapse of thick shells of fluid into a black hole as described in the figure drawn in \cite{Penrose:1964wq} in the isotropic homogeneous universe governed by the Einstein gravity with cosmological constant adopting the interpolating method.
This model has a richer structure than the one studied in \cite{Yokoyama:2021nnw} in the sense that there exists a perfect fluid inside the collapsing or blowing shells of fluid. (See also \cite{Fayos1991MatchingOT,PhysRevD.45.2732}.)
The construction of such a dynamical solution with fluid collapsing into a black hole is indeed nontrivial since it was shown that an emergent curvature singularity is not generally covered by an event horizon and the cosmic censorship could be violated locally or globally in a collapsing process \cite{Eardley:1978tr,Christodoulou:1984mz,PhysRevD.43.1416,Shapiro:1991zza}. (See \cite{Christodoulou:2008nj,Joshi:2012mk} for reviews and further references.)
For reader's convenience, we illustrate the possible local causal structures of a gravitationally collapsing model of fluid considered in this paper by Penrose-Carter diagrams in Fig.~\ref{PC}. 
We consider a situation where there exists a positive cosmological constant, which generates the de Sitter horizon and changes the far asymptotic structure compared to the flat case. Therefore the Penrose-Carter diagrams for the local structures are modified compared to those given in  \cite{Eardley:1978tr}. 
For the purpose of constructing a completely collapsing solution without a naked singularity, the strategy is to set an ansatz for an interpolating metric connecting to a static black hole solution as the final state and determine an unfixed parameter in the ansatz so as to satisfy the Einstein equation.
We determine the orbit of the shells of fluid by the continuity condition for density and the perpendicular component of pressure. 
It turns out that this gravitationally collapsing or blowing model is correlated to the status of the FLRW universe. That is, in this model, a gravitationally collapsing solution happens only when the region inside described by the FLRW metric is under accelerating shrinkage, while a gravitationally blowing one occurs only when the region inside is under accelerating expansion. 
What we call a gravitationally blowing solution is obtained by a time reversal for a gravitationally collapsing one, and thus such a solution describes the process of a classical evaporation of a white hole. We call this solution gravitational evaporation. 

As a special case, we can consider a situation where there does not exist a perfect fluid inside the shell. In this situation, the region inside the shell expands or shrink exponentially, since there exists only cosmological constant or dark energy inside. Then the orbit of the shell is not determined only by the continuity condition. This is natural because the property of the fluid consisting of shells is not determined in this situation. There is room to specify the property of the shells, and, we determine the orbit of the shells of the fluid by imposing an equation of state for fluid at a layer near the surface. 

Finally we compute the energy distribution inside a fixed radius by adopting the manifest covariant definition of energy presented in \cite{Fock:1959,Aoki:2020prb,Aoki:2020nzm}. 
From this energy distribution one can see that the energy of a perfect fluid is converted into that of the shells of fluid in the region where the shells exist. 
Using this definition we easily check that the total energy of the system is independent of the given time evolution and matches the quasi-local mass evaluated at the behavior of the metric in the far asymptotic region. 

The rest of this paper is organized as follows. 
In section \ref{Interpolation} we explain that the Gullstrand-Painlev\'e coordinate system suits to describing gravitational collapse of thick shells of fluid in the isotropic homogeneous universe. 
In section \ref{ShellWithPerfectFluid} we determine the scale factor for a perfect fluid together with cosmological constant and investigate the continuity condition for physical quantities between the region of a perfect fluid and that of shells of fluid, from which the orbit of the shells is determined. 
In section \ref{Energy} we compute the energy distribution of the system and confirm the conservation of the total energy. 
Section \ref{Discussion} is devoted to summary and discussion. 
In appendix \ref{CollapseInflation} we determine the orbit of shells of fluid in the absence of a perfect fluid inside by imposing an equation of state near the surface.

\section{Gullstrand-Painlev\'e coordinates as interpolating metric}
\label{Interpolation}

We investigate the spherically symmetric gravitational collapse into a black hole or the gravitational evaporation from a white hole in the homogeneous isotropic universe, which is described by the FLRW metric with conformally flat space as 
\beal{
ds_{\rm FLRW}^2
=& -(dx^0)^2 +a(x^0)^2({d\check r^2} + \check r^2\tilde g_{ij}dx^idx^j ),
\label{FLRW}
}
where the scale factor $a(x^0)>0$ is only dependent on the time coordinate $x^0$, and $\tilde g_{ij}$ is a metric for the unit $(d-2)$-dimensional Einstein manifold with $\tilde R_{ij}=(d-3)\tilde g_{ij}$.
The behavior of the scale factor is determined by a state of a perfect fluid filled in the region described by the metric. 

In order to construct a solution for gravitational collapse or evaporation in the homogeneous isotropic universe, what we first do is to find a metric interpolating between the FLRW one \eqref{FLRW} and the Kottler spacetime \cite{Kottler1918,Weyl1919} given by 
\beal{ 
ds_{\rm Kott}^2
=& - (1+u(r)) (dx_s^0)^2 +{1\over (1+u(r))} dr^2 + r^2\tilde g_{ij}dx^idx^j , 
\label{Schwarzschild}
}
where $u(r)=-\frac{2\Lambda}{(d-2)(d-1)}r^2 -\frac {m}{r^{d-3}}$ with $m_0$ a positive parameter related to the black hole mass and $\Lambda$ the cosmological constant. 
It was pointed out in \cite{Adler:2005vn} that such an interpolation can be nicely done by the Gullstrand-Painlev\'e coordinates \cite{gullstrand1922allgemeine,1921CR....173..677P}: 
\beal{
ds_{\rm GP}^2
= -(1 -\psi_\pm^2) (dt_\pm)^2 \pm 2\psi_\pm dt_\pm dr + dr^2 + r^2\tilde g_{ij}dx^idx^j , ~~ 
\label{GPpm}
}
where $\psi_\pm$ generally depends on the time and the radial coordinates. We call the coordinate system with the upper sign chosen the upper patch denoted by $U_+$ and the one with the lower sign the lower patch by $U_-$ for convenience. 
These two local patches are not transformed to each other for general $\psi_\pm$, but they are if the two functions have the same absolute value, $\psi_+=\pm\psi_-$, in which case the coordinate transformation is simply $t_+=\mp t_-$ with the double signs in the same order. 
In what follows, for notational simplicity, we write $t=\pm t_\pm$, $\psi=\psi_\pm$, so that 
\beal{
ds_{\rm GP}^2
= -(1 -\psi^2) (dt)^2 + 2\psi dt dr + dr^2 + r^2\tilde g_{ij}dx^idx^j. 
\label{GP}
}
To see the interpolation, we change from the comoving radial coordinate $\check r$ to the proper one by performing a coordinate transformation for the FLRW metric \eqref{FLRW} to absorb the scale factor into the radial coordinate by $r=a(x^0)\check r, t=x^0$ with other coordinates intact. 
Then the above FLRW metric is expressed by the Gullstrand-Painlev\'e coordinates \eqref{GP} with $\psi =- H(t) r$ and $H(t)={d\log a(t) \over dt} $.
Remark that $\psi$ has the opposite sign to $H(t)$, so that $\psi$ is negative for accelerating expansion, while $\psi$ is positive for accelerating shrinkage.
On the other hand, to see the Gullstrand-Painlev\'e coordinates to admit the Kottler black hole metric, we perform a coordinate transformation such that 
\be 
t= x^0_s- g(r), ~~~ 
g(r) = \int dr{-\psi(r) \over 1 + \psi(r)^2},
\ee
where we assume that $\psi$ is a function only of the radial coordinate $r$. 
Then the Gullstrand-Painlev\'e coordinates \eqref{GP} describe the Kottler metric \eqref{Schwarzschild} with $\psi = -\sqrt{-u(r)}$. 
Notice that this transformation is meaningful as long as the function $u(r)$ as well as $\psi$ are nonpositive for all $r>0$. 
Therefore, firstly in order to study gravitational collapse or evaporation described by the Gullstrand-Painlev\'e coordinates, we need to assume that the cosmological constant is nonnegative. This means that there exists a nonnegative constant $H_0$ such that 
\be
H_0^2={2\Lambda \over (d-1)(d-2)}. 
\ee
Secondly, the region of the Kottler black hole and the one described by the FLRW metric cannot be covered by only one of the coordinate patches $U_\pm$, which will be seen in the next section. 

Now that the Gullstrand-Painlev\'e coordinates \eqref{GP} suit to describe a gravitational collapse or evaporation in the FLRW universe, we choose it as the ansatz to solve the Einstein equation. 
By computing the Einstein tensor from the Gullstrand-Painlev\'e coordinates \eqref{GP}, we can determine the form of the stress energy tensor to satisfy the Einstein equation $T_{\mu\nu}= \frac1{8\pi G_N}(R_{\mu\nu}-\half Rg_{\mu\nu} +\Lambda g_{\mu\nu})$, where $G_N$ is the Newton constant:
\beal{
T^t\!_t
=& {d-2 \over 16\pi G_N} { (r^{d-3} (-\psi^2 +H_0^2 r^2 ) )'\over r^{d-2}} , ~~ \nn
T^t\!_r =&{d-2 \over 16\pi G_N}{2\psi \dot\psi \over r} , ~~~~
T^r\!_t =0 , ~~\nn
T^r\!_r =&{d-2 \over 16\pi G_N}\({(r^{d-3} (-\psi^2 +H_0^2 r^2 ) )'\over r^{d-2}}   +\frac{2\dot\psi}{r}\), \nn
T^{i}\!_{j}
=&{\delta^{i}_{j}\over 16\pi G_N} \(
\frac{(r^{d-3} (-\psi^2 +H_0^2 r^2 ))''}{r^{d-3}} +2 \frac{(r^{d-3} \dot\psi)'}{r^{d-3}} \), 
\label{EMT}
}
where $\dot F:= \partial_t F, F':=\partial_r F$. 
We have checked that this stress energy tensor satisfies the covariant conservation equation at $d=4$. 
The stress energy tensor is related to macroscopic observable quantities for the infalling fluid as 
\be 
T^\mu\!_{\nu} = \rho u^\mu u_\nu +q^\mu u_\nu +u^\mu q_\nu + P^\mu\!_\nu, 
\ee
where $u^\mu$ is the fluid velocity relative to a radially moving frame, $\rho$ is the density, $q^\mu, P^\mu\!_\nu$ are a vector and a tensor perpendicular to $u^\mu$, respectively. Note that due to the perpendicularity the tensor $P^\mu\!_\nu$ can be written as 
\be 
P^a\!_b = p_{\perp}(\delta^a\!_b+u^a u_b), ~~ 
P^i\!_j = p_{\parallel}\delta^i\!_j ,  
\ee
where $a,b=t,r$ and $p_\perp, p_\parallel$ are the radial, angular component of the pressure, respectively. 
We fix the radial moving frame so that the fluid velocity be well-defined during the whole dynamical process. 
From a radially moving frame the fluid velocity is written as $(u^\mu)=(1,u^r,\vec0)$, where the time component is normalized to be unity. 
Furthermore we consider a perfect fluid, which is generally timelike, a fluid velocity is normalized so that $u^\mu u_\mu=-1$. From these two equations the radial component is uniquely fixed as $u^r=-\psi$.  
Then the physical quantities read
\bes{
\rho=&-{d-2 \over 16\pi G_N} { (r^{d-3} (-\psi^2 +H_0^2 r^2 ) )'\over r^{d-2}},
\\
p_{\perp}=&{d-2 \over 16\pi G_N} ({(r^{d-3} (-\psi^2 +H_0^2 r^2 ) )'\over r^{d-2}} +\frac{2\dot\psi}{r}),\\
p_{\parallel}=&{1\over 16\pi G_N}  \frac{(r^{d-3} (-\psi^2 +H_0^2 r^2 ))''+2(r^{d-3} \dot\psi)'}{r^{d-3}}, 
\label{rhopp}
}
and $q^\mu=0$.
We shall see that these quantities are indeed well-behaved in later sections. 

Before moving on detailed calculation, let us comment on a radial geodesic on the Gullstrand-Painlev\'e coordinates. 
The particle radial geodesics is determined by $ds^2=-d\tau^2 $ with $\tau$ the particle proper time, which reads 
\be
-(1 -\psi^2) ({dt\over d\tau})^2 +2\psi {dt\over d\tau}{dr\over d\tau}+ ({dr\over d\tau})^2  = -1.
\ee
On the other hand, the time component of the radial geodesic equation reads 
\beal{ 
{d^2t\over d\tau^2} -\psi^2 \psi' ({dt\over d\tau})^2-2\psi \psi' {dt\over d\tau}{dr\over d\tau} - \psi' ({dr\over d\tau})^2 = 0 , 
% ~~  \partial_\tau^2 r + \half \check v^2(-\partial_v f + f\partial_r f) + {dr\over d\tau} \check v \partial_r f = 0
}
where we used $\Gamma^{t}_{tt} =-\psi^2 \psi', ~ \Gamma^{t}_{tr} =-\psi \psi', ~
\Gamma^{t}_{rr} =-\psi'$. Removing ${dr\over d\tau}$ from these two equations we find  
\beal{ 
{d^2t\over d\tau^2} -\psi' ({dt\over d\tau} )^2 = -\psi'. 
}
This equation has an obvious solution of $t = \tau + \const$ 
for any function $\psi$. This means that the time coordinate in the Gullstrand-Painlev\'e metric can be understood as the proper time of a free particle.
Then this reduces the particle radial geodesic equation to
\beal{
{dr \over d t} = - \psi(t,r(t)). 
\label{FreeFall}
}
Plugging $\psi = -H(t)r$ into this gives $\dot r = H(t)r(t)$, which shows that the radial coordinate of a particle geodesic $r(t)$ is proportional to the scale factor $a(t)$. 
This is also expected from the transformation from the FLRW metric to the Gullstrand-Painlev\'e one: $r=a(t) \check r$. 
On the other hand, the radial null geodesic is determined by $ds^2 =0$.
This reads ${dr \over dt} =\mp1 -\psi$. Thus the infinite redshift surface is determined by $\psi =\mp1$. Note that the positive sign corresponds to an outgoing light-ray, while the minus one an ingoing one in the coordinate system \eqref{GP}.

\section{Collapsing shells of fluid with a perfect fluid inside}
\label{ShellWithPerfectFluid} 

\subsection{Scale factor for a perfect fluid with the cosmological constant}
\label{ScaleFactor} 

Before studying gravitational collapse or evaporation, we consider spacetime filled with a perfect fluid and the dark energy or the cosmological constant in general and determine the behavior of its scale factor. 
As argued in the previous section, this situation is realized by setting $\psi=-H(t) r$ in the Gullstrand-Painlev\'e metric \eqref{GP}. 
Indeed in this situation the radial component of the pressure and the angular one determined as \eqref{rhopp} become identical: 
\beal{
p_{\parallel}
=p_\perp={d-2\over 16\pi G_N } (-(d-1)(H(t)^2-H_0^2)- 2\dot H(t) ).
}

Let us consider a perfect fluid to satisfy an equation of state such that 
\beal{ 
p=w\rho , 
\label{EOS}
}
where $p =p_{\parallel}=p_\perp$, $w$ is a constant greater than or equal to $-1$. 
Using \eqref{rhopp} we can rewrite this as 
\beal{
\dot H(t) =- \half (w+1) (d-1) ( H(t)^2 -H_0^2). 
\label{EOS2}
}
Nontrivial solutions are either 
\be 
H(t) =H_0 \tanh\( {H_0\over2} (w+1) (d-1) (t - t_0) \) ~~ {\rm or} ~~ H_0 \coth\(  {H_0\over2} (w+1) (d-1)(t - t_0) \), 
\label{2solutions}
\ee
where $t_0$ is an integration constant. The former solution goes to vanishing in the zero cosmological constant limit $H_0\to0$, which is not the correct behavior of the scale factor. Therefore we choose the latter.  
Then we can determine the scale factor $a(t)$ by solving $\dot a(t)=H(t)a(t)$ as 
\beal{ 
a(t) = C_1 \(\sinh(\pm {H_0\over2} (w+1) (d-1) (t-t_0) )  \)^{2\over (w+1) (d-1) } , 
\label{ScaleFactor}
} 
where $C_1$ is another integration constant and the double sign is chosen so that the term inside the bracket becomes positive. 
This solution correctly reduces to the well-known scale factor for a perfect fluid with the equation of state \eqref{EOS} in the zero cosmological constant limit: 
\beal{
a(t) \to  a_0 |H_0(t-t_0)|^{2\over (w+1)(d-1) },
}
where we set $a_0:=C_1({(w+1)(d-1)\over2})^{2\over (w+1)(d-1)}$. 

Without loss of generality, we can set the integration constant $t_0$ to zero by shifting the time coordinate, so that 
\be 
H(t) = H_0 \coth\(  {H_0\over2} (w+1) (d-1)t \). 
\label{Hubble}
\ee
Then the sign of $H(t)$ is correlated to that of the time coordinate: for $t<0$, $H(t)<0$, or the internal region is shrinking with acceleration,  
while for $t>0$, $H(t)>0$, the internal region is expanding with acceleration. 
In what follows, we use this convention. 

Note that neither solutions of \eqref{2solutions} go to the desired solution of \eqref{EOS2} in the limit $w\to -1$, in which case the correct solution would be such that $H(t)$ is a general constant.
Therefore we have to deal with the case of $w=-1$ or $H(t)=H_0$ separately when we construct a solution describing gravitational collapse or evaporation in this case.  
In this situation we cannot determine the orbit of the shell only by the continuity condition. One of the ways to determine the orbit of the shell is to specify the property of the fluid by imposing an equation of state near the surface of the shell \cite{Yokoyama:2021nnw}. The analysis becomes more complicated. For completeness we perform it in appendix \ref{CollapseInflation}.

\subsection{Shell structure and continuity condition} 
\label{ContinuityCondition} 

Let us construct a solution for gravitational collapse or evaporation of shells of fluid in the presence of a perfect fluid inside and the cosmological constant filled in the spacetime.
In order to construct such a solution of our interest, we first set the ansatz to realize such a configuration that a thick shell of fluid exists on top of a perfect fluid.
Thus at a fixed time slice the space is generally separated into three regions denoted by I, II, III as explained in the introduction: the region I is filled with a uniform perfect fluid, the region II is with thick shell of fluid infalling, and the third one is only with dark energy or cosmological constant.  
To describe the regions precisely we prepare for a function $h(r)$ to describe the radial orbit of the thick shell of fluid such that the orbits of the two edges of the thick shell are described by $t=h(r)$ and $t=h(r)+\Delta$, where $\Delta$ is a positive constant describing a width of the shells.\footnote{We flip the sign of the function $h(r)$ in this paper compared to the one in \cite{Yokoyama:2021nnw}. }
More precisely, if $h(r)$ is a monotonically decreasing function, in which case gravitational collapse happens, each region is described as follows: 
\beal{
\left\{
\begin{array}{rcl}
 \mbox{region I} &=& \{(t,r); 0\leq r\leq h^{-1}(t) \} \\
 \mbox{region II} &=& \{(t,r); h^{-1}(t)\leq r\leq h^{-1}(t-\Delta) \} \\
 \mbox{region III} &=& \{(t,r); h^{-1}(t-\Delta)\leq r \} \\
\end{array}
\right.  
\quad \mbox{if $h'(r)<0$.}
}
On the other hand, if $h(r)$ monotonically increases, then gravitational evaporation happens, and each region is described as  
\beal{
\left\{
\begin{array}{rcl}
 \mbox{region I} &=& \{(t,r); 0\leq r\leq h^{-1}(t-\Delta)\} \\
 \mbox{region II} &=& \{(t,r); h^{-1}(t-\Delta)\leq r\leq h^{-1}(t) \} \\
 \mbox{region III} &=& \{(t,r);h^{-1}(t)\leq r \} \\
\end{array}
\right.  
\quad \mbox{if $h'(r)>0$.}
}
A goal of the analysis is to determine the function $h(r)$ to satisfy the Einstein equation with some physical condition such as the continuity at interfaces and an equation of state of fluid. 
Note that the orbit of a particle which consists of the shells satisfies a differential equation such that 
\be 
{dr \over dt} = \frac1{h'(r(t))}.
\label{ParticleOrbit}
\ee

Let us take a strategy so as to consider solutions in the region I and III and connect them to the region II smoothly. 
Without losing generality we can use the coordinate system with the upper patch $U_+$ to cover the region I. As discussed in the previous section, the Einstein equation can be solved in the region I with the function $\psi_+$ chosen as
\be 
\psi_+=-H(t) r, \quad \mbox{$(t,r) \in$ region I},
\label{regionI+}
\ee
where $H(t)$ is given by \eqref{Hubble}. 
Therefore if $t<0$ then $\psi_+ \geq0$, while $\psi_+ \leq0$ if $t>0$, in this region. 
On the other hand, since the metric describes the Schwarzschild black hole in the region III, as argued in the previous section, $\psi_+$ may be written as     
\be 
\tilde\psi_+ \overset?=-\sqrt{ {m_0\over r^{d-3}} +H_0^2 r^2 }, \quad \mbox{$(t,r) \in$ region III}.
\label{regionIII+}
\ee
Let us try to connect these solutions via the region II. 
A candidate of a desired function may have a form such that   
\be 
\tilde\psi_+ \overset?= -\sqrt{ {m(t, r)\over r^{d-3}} +H(t,r)^2 r^2 }, \quad \mbox{$(t,r) \in$ region II},
\ee
where $m(t,r), H(t,r)$ are smooth functions satisfying 
\bes{ 
\lim_{(t,r)\to \rm region\,I} m(t, r) = 0, ~~
\lim_{(t,r)\to \rm region\,I} H(t, r) = H(t) ,\\
\lim_{(t,r)\to \rm region\,III} m(t, r) = m_0, ~~
\lim_{(t,r)\to \rm region\,III} H(t, r) = H_0.
\label{smooth} 
}
However, it is soon noticed that this function cannot be smooth at the interface between the region I and II, since $H(t)<0$ for $t<0$, so that $\lim_{(t,r)\to \rm region\,I}\tilde\psi_+(t,r)= H(t)r \not=\psi_+$ for $t<0$. 

This shows that it is not possible to cover all the regions I, II and III by one local patch $U_+$. 
Thus we use the local upper patch $U_+$ to cover the region I and II and set the function $\psi_+$ in the region II as 
\be 
\psi_+(t_+, r) = \sqrt{ {m(t_+, r)\over r^{d-3}} +H(t_+,r)^2 r^2 }, \quad \mbox{$(t_+,r) \in$ region I and II},
\label{psiansatz+}
\ee
where $m(t,r), H(t,r)$ satisfy \eqref{smooth}, while we cover the region II and III by the local patch $U_-$, where $\psi_-$ is chosen as 
\be 
\psi_-(-t_-, r) = \sqrt{ {m(-t_-, r)\over r^{d-3}} +H(-t_-,r)^2 r^2 }, \quad \mbox{$(-t_-,r) \in$ region II and III}. 
\label{psiansatz-}
\ee
Then $\psi_+$ with \eqref{psiansatz+} smoothly connects to the one with \eqref{regionI+} when there exists $t_+<0$ in the region I, and the de-Sitter Schwarzschild black hole solution is realized in the region III covered by the lower patch $U_-$. 
These two patches can be glued consistently in the intersection by the coordinate transformation $t_+=-t_-$ since $\psi_+ = \psi_-$ in $U_+\cap U_-$. 
In order to argue for the case with $t_+>0$ present in the region I, we consider one global patch $U=U_+\cup U_-$ and cover all the regions I, II and III by $U$, on which the metric is given by \eqref{GP} with  
\be 
\psi(t, r) = \sqrt{ {m(t, r)\over r^{d-3}} +H(t,r)^2 r^2 }. 
\label{psiansatz}
\ee
Then we do not need to care about the issue how to cover each region by local patches. That is, it is sufficient to construct a smooth function $\psi$ for all the regions to satisfy the Einstein equation, and have only to choose an appropriate patch for the physical interpretation. 

The function $m(t,r)$ describes a profile of thick shell of fluid and $H(t,r)$ a profile of a perfect fluid inside it. 
Thus we realize these functions by preparing the corresponding slope functions $F(x)$, $\Theta(x)$, which are monotonically increasing or decreasing functions valued between $0$ and $1$ in the interval $[0,1]$ as 
\beal{
m(t, r)=& m_0 \theta(r) F({t -h(r)  \over \Delta}), 
\label{m}\\
H(t,r)^2=&H_0^2 +\delta H(t)^2\Theta({t -h(r)  \over \Delta}) ,
\label{H}
}
with $\theta(r)$ the step function with $\theta(0)=0$, 
\be 
\delta H(t)^2=H(t)^2 -H_0^2 
=\left\{ 
\begin{array}{lr} 
\frac{H_0^2}{ \sinh^2 ( {H_0 \over 2} (w+1)(d-1) t  ) } & w\not=-1\\
0 & w=-1 
\end{array}
\right. .
\label{deltaH}
\ee 
More specifically whether slope functions $F(x),\Theta(x)$ are upslope or downslope depends on a situation such that 
\be 
\left\{ 
\begin{array}{lc} 
F'(x)\geq0, \Theta'(x)\leq0 & \mbox{with $h'(r)<0$ or gravitational collapse}\\ 
F'(x)\leq0, \Theta'(x)\geq0 & \mbox{with $h'(r)>0$ or gravitational evaporation}. 
\end{array}
\right. 
\label{sigma}
\ee
The condition \eqref{smooth} is satisfied 
$F(0)= \Theta(1)=\sigma,  F(1)= \Theta(0)=1-\sigma $, where $\sigma$ is defined by 
\be 
\sigma:=
\left\{ 
\begin{array}{lc} 
0, & \mbox{with $h'(r)<0$ or gravitational collapse}\\ 
1, & \mbox{with $h'(r)>0$ or gravitational evaporation}
\end{array}
\right. 
\label{sigma}
\ee
More formally, $\sigma=\theta(h'(r))$. 
% Let us denote $F(x):= F(x),~~ \Theta(x):=\Theta(x)$ with $0\leq x\leq1$ for later convenience. 

These slope functions are inserted so as for some physical quantities to be continuous at the interfaces. 
We request the density and the perpendicular or radial component of the pressure to be continuous. 
We first consider the continuity condition at interface between the region I and II. 
The continuity of the density at this interface is mathematized as  
\be 
\lim_{ x\to \sigma+0}\rho|_{t=h(r)+\Delta x}=\lim_{ x\to \sigma-0}\rho|_{t=h(r)+\Delta x},
\ee
The density is given in each region as 
\be 
\rho =
\left\{
\begin{array}{lr}
 {(d-2)\delta H(t)^2 (d-1)\over 16\pi G_N }, & \mbox{region I}\\
  {(d-2)(m_0F'({t -h(r) \over \Delta}){-h'(r) \over \Delta}  +\delta H(t)^2 \{\Theta'({t -h(r) \over \Delta}){-h'(r) \over \Delta}  r^{d-1} + \Theta({t -h(r) \over \Delta})(d-1)r^{d-2} \} )\over 16\pi G_N r^{d-2}}, & \mbox{region II} \\
 0, & \mbox{region III}\\
\end{array}
\right.
\label{rho}
\ee
Therefore the continuity condition reads 
\beal{
\lim_{ x\to\sigma} [
m_0F'( x) +\delta H(t)^2 \Theta'( x) r^{d-1}] = 0. 
\label{continuitydensity}
} 
For the case with $w=-1$, $\delta H(t)$ vanishes. Thus this condition is equivalent to $F'(\sigma)=0$. 
For the case with $w\not=-1$, the value 
\be 
\alpha:= -\lim_{ x\to\sigma} \frac{\Theta'( x)}{F'( x)} 
\ee
is nonnegative since the slope functions $F, \Theta$ have the opposite sign of the gradient. We assume $\alpha$ to be nonzero. 
Then employing \eqref{deltaH} we obtain 
\beal{
\sinh^2 ( {H_0 \over 2} (w+1)(d-1)(h(r) +\sigma\Delta) ) = {H_0^2 \alpha r^{d-1} \over m_0 }.
}
This can be solved as $h(r) +\sigma\Delta = \pm \frac2{{H_0 } (w+1)(d-1)}\arcsinh(\sqrt{{\alpha \over m_0} } H_0r^{d-1\over2}),$ which can be rewritten as 
\beal{
h(r) 
=\left\{
\begin{array}{lr}
- \frac2{{H_0 } (w+1)(d-1)}\arcsinh(\sqrt{{\alpha \over m_0} } H_0r^{d-1\over2}) , & h'(r)<0 \\
-\Delta + \frac2{{H_0 } (w+1)(d-1)}\arcsinh(\sqrt{{\alpha \over m_0} } H_0r^{d-1\over2} ) , &h'(r)>0 \\
\end{array}
\right. . 
\label{hsolution}
}
Comments are in order.\footnote{ A similar expression was obtained as (40) in \cite{Cissoko:1998mx} from the junction condition. Just for comparison, we take the result in \cite{Markovic:1999di}, where a collapsing pressureless fluid (or dust) with density finely tuned as $\rho=\mu/a^3$ was studied. The equation of the trajectory of the surface is determined as the equation (40), which reads ${dt \over dr}=\frac{-1}{(1-(H_0^2 r^2+\frac{m_0}r))\sqrt{H_0^2 r^2+\frac{m_0}r}}$ in our notation. On the other hand, in our study of a collapsing fluid with a perfect fluid inside, the corresponding quantity is ${dh \over dr}=\frac{-1}{(1+w) \sqrt{H_0^2 r^2+{m_0 \over \alpha}\frac1r}}$.}
Firstly, the gravitational collapsing/evaporating solution is realized accordingly when the FLRW universe shrinks/expands respectively in this model.
Secondly, in the zero cosmological constant limit, this solution reduces to   
\beal{
h(r) 
\to\left\{
\begin{array}{lr}
- \frac2{(w+1)(d-1)}(\sqrt{{\alpha \over m_0} } r^{d-1\over2}) , & h'(r)<0 \\
-\Delta + \frac2{ (w+1)(d-1)}(\sqrt{{\alpha \over m_0} } r^{d-1\over2} ) , &h'(r)>0 \\
\end{array}
\right. .
\label{hsolutionLambda0}
}
The dependence of the radial coordinate matches the result in \cite{Adler:2005vn} at $d=4$. 
We shall show soon that the emergent curvature singularity is always covered by the emergent event horizon. 
On the other hand, on the continuity of the radial component of the pressure, which reads $p_\perp= - \rho + {d-2 \over 16\pi G_N}{2\dot \psi(t,r) \over r}$, it is necessary and sufficient to consider the continuity for $\dot \psi$ taking into account the continuity of $\rho$. $\dot \psi$ is given by 
\be 
\dot\psi=
\left\{
\begin{array}{lr}
 \frac{1 }{2\psi }( \partial_t \delta H(t)^2r^2), & \mbox{region I}\\
 \frac{1}{2\psi }\( {m_0F'({t -h(r) \over \Delta}){1 \over \Delta} \over r^{d-3}} +\{ (\partial_t \delta H(t)^2 )\Theta({t -h(r) \over \Delta})+ \delta H(t)^2 \Theta'({t -h(r) \over \Delta}) \frac1\Delta \} r^2 \), & \mbox{region II} \\
 0, & \mbox{region III}\\
\end{array}
\right.
\label{dotpsi}
\ee
Thus the continuity condition, $\lim_{ x\to \sigma+0}\dot\psi|_{t=h(r)+\Delta x}=\lim_{ x\to \sigma-0}\dot\psi|_{t=h(r)+\Delta x}
$, reduces to the same one for the density $\rho$, \eqref{continuitydensity}. 
As a result the orbit of the shell of fluid is determined and both gravitational collapse and evaporation are possible. 

We next consider the continuity condition at the interface between II and III. 
That of the density is now written as 
\be 
\lim_{ x\to \sigma+0}\rho|_{t=h(r)+\Delta- \Delta x}=\lim_{ x\to \sigma-0}\rho|_{t=h(r)+\Delta-\Delta x}. 
\ee
From \eqref{rho} the continuity condition for the density reads 
\beal{
m_0F'(1-\sigma){-h'(r) \over \Delta}  +\delta H(t)^2 \(\Theta'(1-\sigma){-h'(r) \over \Delta}  r^{d-1} + \Theta(1-\sigma)(d-1)r^{d-2} \) =0.
\label{continuityrhoII}
}
For the case with $w=-1$, $\delta H(t)=0$, which reduces this condition to $F'(1-\sigma)=0$. 
For the case with $w\not=-1$, from the continuity condition between the region I and II, $h(r)$ is determined as \eqref{hsolution}. 
Combining this, the above condition \eqref{continuityrhoII} is equivalent to 
\beal{ 
F'(1-\sigma)=\Theta'(1-\sigma)= 0. 
\label{continuityII}
}
As in the same computation of the continuity condition between the region I and II, 
we can show that the continuity of the perpendicular component of the pressure reduces to the same continuity condition for the density \eqref{continuityII}. 

We conclude that the function $h(r)$ is determined as \eqref{hsolution} from the continuity condition for $w\not=-1$, in which there exists a non-trivial perfect fluid inside. Gravitational collapse/evaporation happens when the internal FLRW universe shrinks/expands, respectively. 
On the other hand, the function $h(r)$ cannot be determined only from the continuity condition in the case with $w=-1$, where the inside of the shell of fluid is filled only with the dark energy. 
This implies that one needs to impose a condition on thick shell of fluid to determine its orbit in this situation, which is physically reasonable since the property of fluid is not specified in this case. 
We determine the function $h(r)$ in this case by imposing an equation of state near the interface in the appendix. 

\begin{figure}
  \begin{center}
  \subfigure[Blackhole formation]{
	\includegraphics[scale=.4]{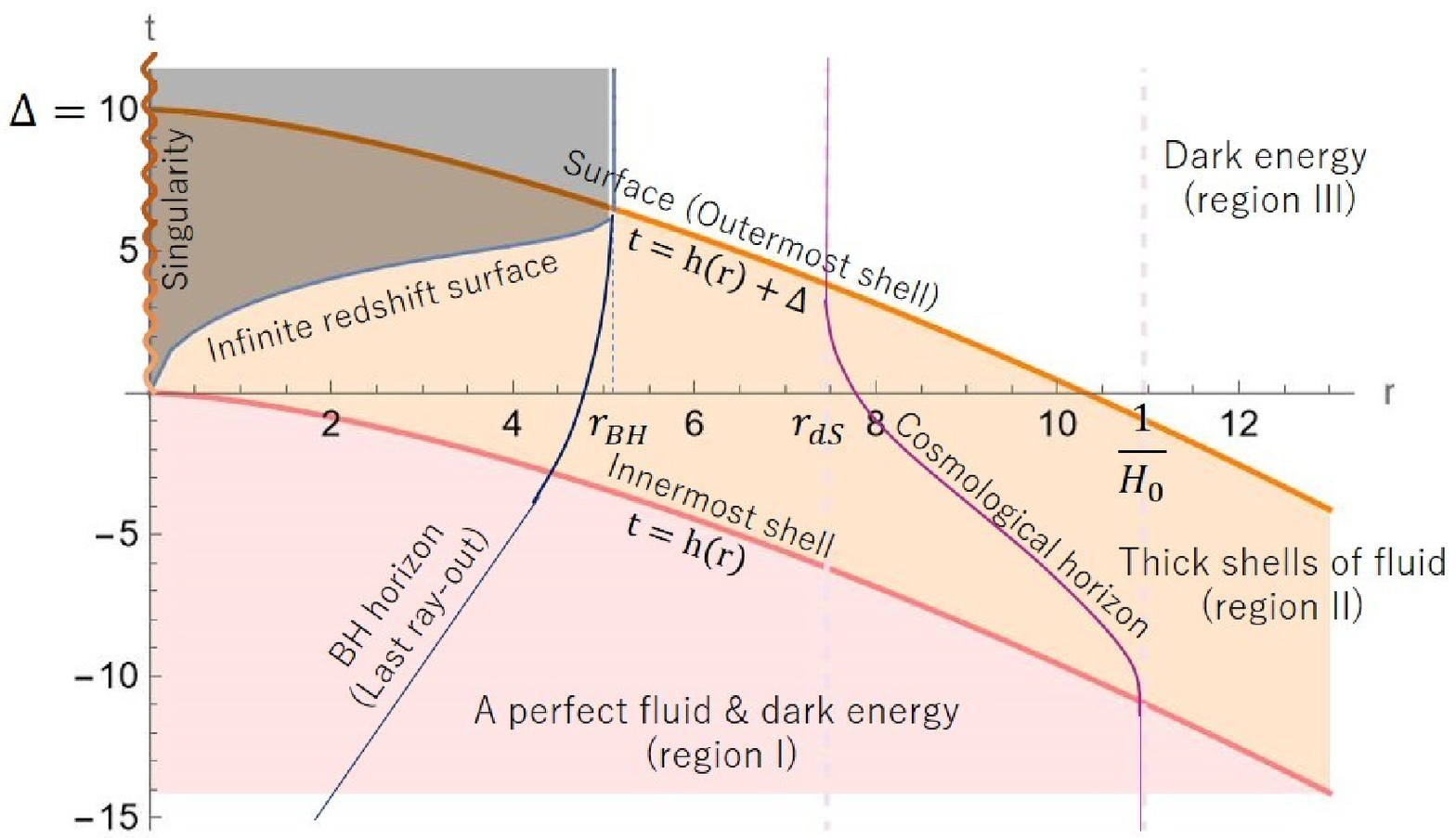}
  }
  \qquad\qquad
  \subfigure[Whitehole evaporation]{
	\includegraphics[scale=.4]{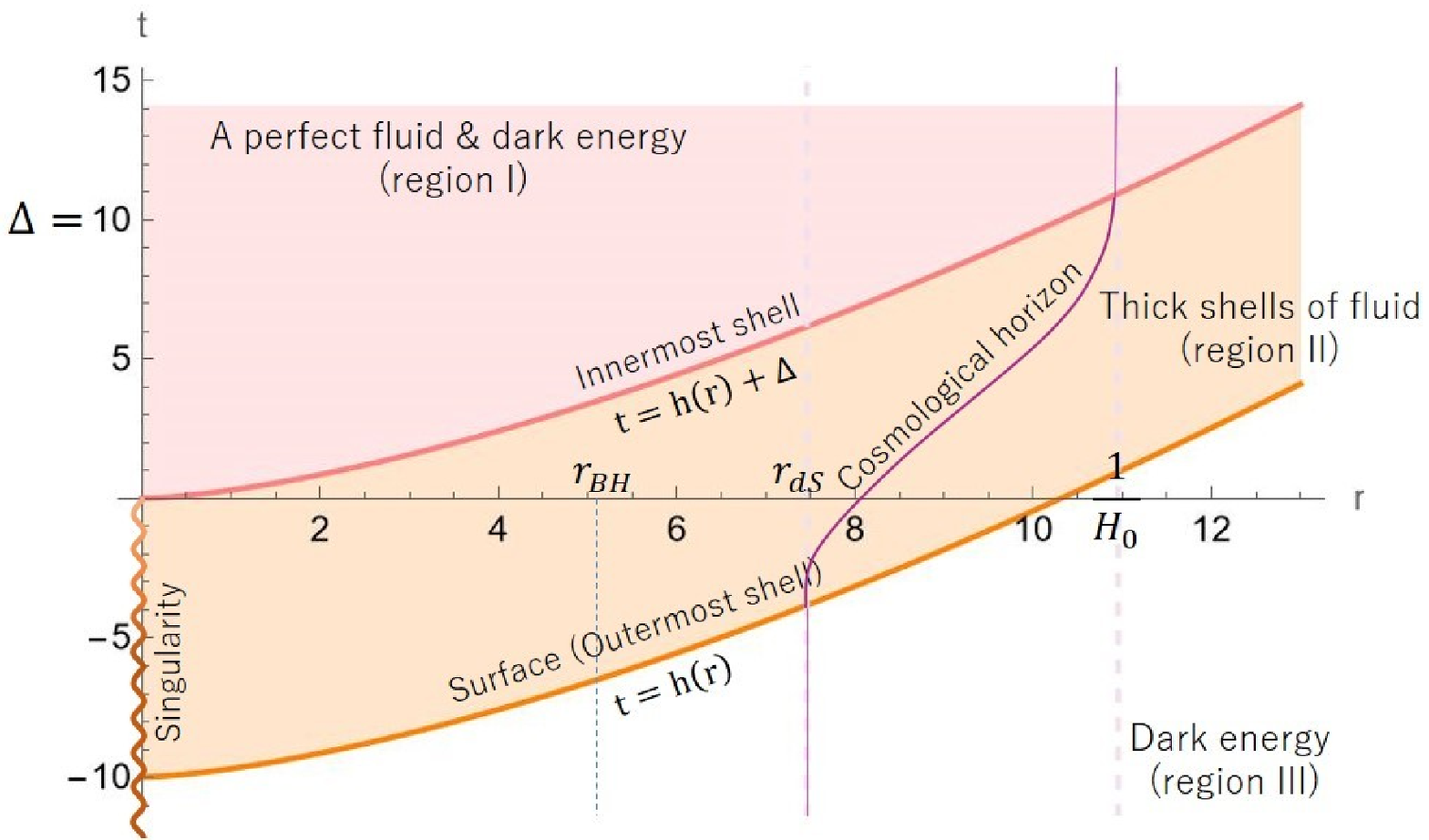}
  }
  \end{center}\vspace{-0.5cm}
  \caption{(a) depicts the blackhole formation from infalling shells of fluid. The bottom one colored by pink is the region I, where a perfect fluid and the dark energy coexist, the middle one colored by orange is the region II, where the shells of fluid are infalling, and the upper one is the region III, where there exists only the dark energy.
  The orbit of the thick shells of fluid is drawn by choosing the parameters $w=0.1, \Delta=10, m_0=4, \Lambda=0.025$, $F(x)=\sin^2{\pi x\over 2}, \Theta(x)=\cos^2{\pi x\over 2}$. The innermost shell reaches the origin at $t=0$, at which the curvature singularity starts to emerge. The growing infinite redshift surface is given by $t= h(r) + \Delta F^{-1}(\frac rm(1-\frac\Lambda3 r^2))$. 
  Other curves such as the last ray-out, the location of the cosmological horizon, which are described by some differential equations explained in the main text, are schematically drawn and not precise. 
  (b) depicts the whitehole evaporation into blowing shells of fluid. We chose the same values for the parameters. 
  }
  \label{Fig:D10m4}
\end{figure}

For instruction we draw a schematic picture for the case of gravitational collapse of this model in four dimensions, $d=4$, in Fig.~\ref{Fig:D10m4}.
The location of the event horizon of the final black hole is determined by $\psi=1$, where $\psi$ takes the form in the region III: $\psi=\sqrt{\frac{m_0}r + r^2H_0^2}$. 
This equation is solved by the Cardano's formula as $r=r_k$, where 
\be 
r_k =\omega^k \sqrt[3]{-\frac{m_0}{2H_0^2}+\sqrt{(\frac{m_0}{2H_0^2})^2-(\frac 1{3H_0^2})^3}}+\omega^{3-k}\sqrt[3]{-\frac{m_0}{2H_0^2}-\sqrt{(\frac{m_0}{2H_0^2})^2-(\frac 1{3H_0^2})^3}}, ~~~~~
\ee
with $k=0,1,2$, and $\omega={-1+i\sqrt3\over2}$ the cube root of unity. 
As long as the $H_0$ is smaller than the critical value $\frac2{3\sqrt3m_0}$, there exist 2 positive solutions out of 3 and the smaller one is the radius of the location of the final black hole horizon, $r_{\rm BH}=r_1$, while the bigger one is that of the location of the de-Sitter horizon, $r_{\rm dS}=r_2$. 
On the other hand, the growing infinite redshift surface is determined by $\psi=1$ in the region II, where $\psi$ is given by $\psi=\sqrt{ \frac{m_0F(\frac{t-h(r)}\Delta)}r + r^2H_0^2(1+\frac{\Theta(\frac{t-h(r)}\Delta)}{\sinh^2(\frac{H_0}2 (w+1)3t)}) }$. 
The infinite redshift surface extends up to the location of the final black hole horizon, which intersect at the time $t_{BH}=\Delta+h(r_{BH})$. For later convenience we denote the intersecting point $(t_{BH},r_{BH})$ by $P$. 
Note that the explicit expression of the infinite redshift surface depends on a choice of the slope functions $F, \Theta$. 

Now let us show that the emergent curvature singularity is covered by the emergent event horizon. This is sufficient to show that there exists a light-ray which is emitted outward from the origin before the singularity emerges but does not go far away to infinity. 
A light-ray emitted outward from the origin is described by the radial null geodesic  ${dt\over dr} =1-\psi$. For a while after the emittance, the light-ray moves in the perfect fluid or the region I. Therefore during this period $\psi = -H(t)r$, where $H(t)$ is given by \eqref{Hubble}. Note that during this period, $t<0$, and thus $H(t)<0$. Then after some time the right-ray reaches the interface between the perfect fluid and the thick shell, and starts to go through the thick shell or the region II. During this period, $\psi =\sqrt{ {m(t, r)\over r^{d-3}} +H(t,r)^2 r^2 }$. 
This holds for all the right-rays before they get out of the region II. 
When the right-lay comes out the region II, whether it goes outward or inward depends on when it is emitted. 
The light-ray which passes through the point $P$ does not come out, because it is trapped at the event horizon of the final static black hole geometry. This is the last light-ray and its orbit yields that of the event horizon of the growing black hole. 
Any light-ray emitted before the last light-ray goes outward to infinity, while any one  emitted after the last light-ray goes inward and is trapped inside. 
Thus the emergent curvature singularity in this model is always covered by the event horizon. 

\section{Energy distribution and conservation}
\label{Energy}

We compute the energy distribution of the system using the definition given in \cite{Fock:1959,Aoki:2020prb,Aoki:2020nzm}. 
We denote the energy inside the radius $R$ at time $t$ by $E(t,R)$:
\beal{
E(t,R)
=&\int_{\sqrt{x^2}\leq R} d^{d-1} x\sqrt {|g|} T^t{}_\mu n^\mu, 
\label{EnergyR}
}
where $n^\mu = -\delta^\mu_t$.
Using \eqref{EMT} we can compute this as follows. 
\beal{
E(t,R)
=&{d-2 \over 16\pi G_N} V_{d-2} \int_0^R dr {\partial_r  (m(t, r)+(H(t,r)^2 -H_0^2) r^{d-1} )} \nn
=:& E_1(t,R) + E_2(t,R) +E_3 (t,R)
}
where $V_{d-2}$ is the volume of the internal Einstein manifold, 
\beal{
E_1 (t,R)
=& M \int_0^R dr  ({d\theta(r)\over dr}\wt F({t -h(r)\over \Delta}) ) 
= M \wt F({t-h(0)\over \Delta}),  \\
E_2 (t,R)
=& M \int_0^R dr  (\partial_r \wt F({t -h(r)\over \Delta}) ) 
= M( \wt F({t -h(R)\over\Delta})  - \wt F(\frac{t-h(0)}\Delta )) , \\
E_3 (t,R)
=& M \frac{( H(t)^2-H_0^2)}{m_0}\wt\Theta({t -h(R)\over \Delta})R^{d-1} ,
}
where $M={d-2 \over 16\pi G_N} V_{d-2} m_0$ and $\wt F, \wt \Theta$ are extended functions of $F,\Theta$ to the region outside the interval $[0,1]$ with the values unchanged:
\be 
\wt F(x)=
\left\{
\begin{array}{lc}
F(0), & x\leq 0 \\
F(x), & 0\leq x \leq 1 \\
F(1), & 1\leq x\\
\end{array}
\right., \quad 
\wt \Theta(x)=
\left\{
\begin{array}{lc}
\Theta(0), & x\leq 0 \\
\Theta(x), & 0\leq x \leq 1 \\
\Theta(1), & 1\leq x\\
\end{array}
\right.
\ee
$E_1(t,R)$ is the energy localized at the origin or the mass of the black hole at a fixed time $t$. $E_2(t,R)+E_3(t,R)$ is the energy of both the perfect fluid and the shell of fluid inside the radius $R$.
Note that the energy of the perfect fluid inside the shell is converted into that of the shells of fluid in the region II.  

We comment on the conservation of the total energy, which is defined by $E(t)= \lim_{R\to\infty} E(t,R)$.\footnote{The momentum for the radial direction is defined by $P^r=\int_{\mbf R^{d-1}} d^{d-1} x\sqrt {|g|} T^t{}_\mu v^\mu$, where $v^\mu= \delta^\mu_r$.}   
Since this system does not radiate anything, the total energy is expected to be conserved. 
Whether the total energy is conserved or not can be tested by the conservation test equation \cite{Aoki:2020nzm}. 
The statement is that the charge defined by $Q = \int_{\mbf R^{d-1}} d^{d-1} x\sqrt {|g|} T^t{}_\mu v^\mu,$ is conserved if the following equation is satisfied: 
\be 
T^\mu{}_\nu \nabla_\mu v^\nu = 0.   
\ee
To discuss the conservation of the energy, we choose $v^\mu = n^\mu$. 
Since $\Gamma^i_{\mu t}=0$ for all $\mu$, this equation reduces to 
\be 
T^a{}_b \nabla_a n^b = 0 
\ee
where $a,b = t,r$. One can check that this conservation test equation holds by using
\be 
\nabla_a n^b = \left(
\begin{array}{cc}
 \psi^2 \psi' & \psi \psi' \\
 -\psi' \psi^3+\psi' \psi-\dot\psi & -\psi^2 \psi' \\
\end{array}
\right). 
\ee
It is also not difficult to compute the total energy directly by using $\wt F({t -h(\infty)\over\Delta}) = 1, \wt \Theta({t -h(\infty)\over\Delta}) = 0$.
The result is $E(t)=M$, where $M$ is the final or initial black hole mass in the case of gravitational collapse or evaporation, respectively.
The total energy of the system is conserved as expected, and matches the ADM or quasi-local mass computed in the asymptotic behavior of the metric. 

\section{Discussion}
\label{Discussion}
%\noindent
%{\em 4. discussion} \hskip 0.3cm

We have investigated gravitational collapse or evaporation of shells of fluid in the isotropic homogeneous universe without any radiation described by the Einstein gravity with cosmological constant. 
To this end we have chosen the ansatz of a metric written by the Gullstrand-Painlev\'e coordinates which suits to describe the isotropic homogeneous space inside the shells of fluid and the de-Sitter black hole outside them. 
After determining the scale factor for a perfect fluid to obey an equation of state together with cosmological constant, we have determined the orbit of thick shell of fluid from the continuity conditions for the density and the perpendicular component of the pressure at interface. 
When there is no perfect fluid inside the shells, the orbit of the shells cannot be determined by the continuity condition. 
We have determined the orbit of the shells by imposing an equation of state near its surface as done in \cite{Yokoyama:2021nnw} in appendix \ref{CollapseInflation}. 
We have confirmed that the total energy of this model is independent of the given time evolution. 

In this paper we dealt with the matter constituents of shells as fluid.
This approximation can be justified only when the gravitational coupling constant is small enough to neglect the back-reaction. 
If we specified the matter field and took into account back-reaction, then the analysis would be more complicated. 
It would be interesting to study the system specifying the matter and taking into account the back-reaction perturbatively and what happens when the coupling constant is not small. In such a situation the shell crossing singularity would become strong obstruction to constructing a solution as pointed out in \cite{Yokoyama:2021nnw}. 

Another way to make the system closer to a realistic situation is to consider the time dependence of the radial component of the Gullstrand-Painlev\'e metric as considered in \cite{Misner:1964je}. Such a metric can describe not only the radiation of gravitational wave but also the isotropic homogeneous universe with different topology. It would be interesting to evaluate the energy of the gravitational wave by computing the differences of energy between the shells and the final black hole. 
It might be amusing to study the possibility to construct a solution of the Einstein gravity with a change of topology of the universe employing such a generalized Gullstrand-Painlev\'e metric.

We hope to come back to these issues in near future. 

\section*{Acknowledgement}
%\noindent
%{\em Acknowledgement}\hskip 0.3cm
The author would like to thank Shigeki Sugimoto for valuable comments on the draft. 
This work is supported in part by the Grant-in-Aid of the Japanese Ministry of Education, Sciences and Technology, Sports and Culture (MEXT) for Scientific Research (No.~JP19K03847, 19H01897). 

\appendix
\section{White hole evaporation during inflation}
\label{CollapseInflation} 

In this appendix, as asserted in section \ref{ContinuityCondition}, we study the case with $w=-1$ or $H(t)=H_0$, where the internal of the shell is filled only with dark energy, so that the function $\psi$ in the Gullstrand-Painlev\'e metric is given by
\be 
\psi =- \sqrt{ {m(t,r)\over r^{d-3}} +H_0^2 r^2 }, 
\label{psiansatz2}
\ee
where $m(t,r)$ takes the form of \eqref{m}. 
In this case only the continuity condition does not determine the function of the orbit of the shells $h(r)$. 
We determine this function by imposing an equation of state near the surface. 
An equation of state studied in \cite{Yokoyama:2021nnw} is the usual one such that pressure is proportional to density unless the pressure is mixed with its components $p_{\parallel}$ and $p_{\perp}$:
\be 
P =w_0 \rho,
\label{EOSgeneral}
\ee
where $w_0\geq-1$ and $P$ is some mixture of the components of the pressure so that $P = p_{\parallel} + v_0 p_{\perp}$ with $v_0$ a constant. 
Recall that the surface or the outermost shell is specified by $t= h(r) + \Delta(1-\sigma)$, where $\sigma$ is given by \eqref{sigma} and takes $0$ or $1$ for the case of gravitational collapse or evaporation, respectively. 
We impose the equation of state \eqref{EOSgeneral} at a layer specified by $t= h(r) + \Delta(1-\sigma - \varepsilon)$, where $\varepsilon$ is a small positive number. 

\subsection{Transverse pressure}
\label{Transverse} 
We first study the case where the pressure in \eqref{EOSgeneral} consists only of the parallel component or $v_0=0$. In this situation $p_{\parallel}=w^{(0)}_{\parallel}\rho $ with $w^{(0)}_{\parallel} \geq-1$. 
Using \eqref{rhopp} one can rewrite this as 
\be
(-m'(t,r)+ {\dot m(t,r) \over \psi})' =w_{\parallel}\frac1r m'(t,r) , 
\ee
where $w_{\parallel}=(d-2)w^{(0)}_{\parallel}$. 
This can be further computed by employing \eqref{m} as 
\beal{
-yh' (h'  + \frac{1}{\bar\psi}) + (h''  - \frac{{\bar\psi}'}{{\bar\psi}^2})  =-\frac {w_{\parallel}}rh' ,
\label{transverse}
}
where $y= {F''(1-\sigma-\varepsilon) \over F'(1-\sigma-\varepsilon)\Delta}$, ${\bar\psi} =- \sqrt{ {\bar m\over r^{d-3}} +H_0^2 r^2 }$ with $\bar m = m_0 F(1-\sigma-\varepsilon)$ . 

We cannot solve the differential equation \eqref{transverse} analytically with general parameterization, so we solve it simplifying the situation with the help of {\it mathematica}.
We first consider the case with $y=0$.  
In this situation \eqref{transverse} reduces to 
\beal{
h''  - \frac{{\bar\psi}'}{{\bar\psi}^2}  =-\frac {w_{\parallel}}rh', 
}
This can be solved in terms of $h'(r)$ as  
\beal{
h'(r) |_{y=0}
=& c_1 r^{-w_{\parallel}}-r^{\frac{d-3}{2}}\left( \frac{ 2 w_{\parallel} \, _2F_1\left(\frac{1}{2},\frac{d+2 w_{\parallel}-3}{2 (d-1)};\frac{-3 d-2 w_{\parallel}+5}{2-2 d};-\frac{H_0^2 r^{d-1}}{\bar m}\right) }{ (d+2 w_{\parallel}-3) \sqrt{\bar m}}- \frac{1}{ \sqrt{H_0^2 r^{d-1}+\bar m}}\right), \notag
}
where $c_1$ is an integration constant and $_2F_1$ is the hypergeometric function. 
This can be further integrated as 
\beal{
h(r)|_{y=0}
=&\frac{c_1 r^{1-w_{\parallel}}}{1-w_{\parallel}}+\frac{2 \arcsinh\left(\frac{H_0 }{\sqrt{\bar m}}r^{\frac{d-1}{2}}\right)}{H_0(d-1)} \nn
&-\frac{2 w_{\parallel} r^{\frac{d-1}{2}} \Gamma \left(\frac{d+2 w_{\parallel}-3}{2 (d-1)}\right) \, _3{F}_2\left(\frac{1}{2},\frac{d+2 w_{\parallel}-3}{2 (d-1)},\frac{1}{2};\frac{-3 d-2 w_{\parallel}+5}{2-2 d},\frac{3}{2};-\frac{H_0^2 r^{d-1}}{\bar m}\right)}{(d-1)^2 \sqrt{\bar m}\Gamma \left(\frac{d+2 w_{\parallel}-5}{2 (d-1)}\right)}, 
} 
where $_pF_q$ is the generalized hypergeometric function. The last term appears due to the effect of pressure which a particle consisting of fluid receives. 
We fix the integration constant $c_1$ by requesting that the particle radial geodesic reduces to that of a freely falling one in the pressureless limit $w_{\parallel}\to0$. In this limit the solution reduces to 
\beal{
h(r)|_{y=0}\to& c_1 r +\frac{2 \arcsinh\left(\frac{H_0}{\sqrt{\bar m}} r^{\frac{d-1}{2}}\right)}{(d-1) H_0}, 
}
which satisfies $h'|_{y=w_\parallel=0} = c_1-\frac1{\bar\psi}$. Then the particle orbit \eqref{ParticleOrbit} becomes ${dr\over dt} ={1 \over c_1-\frac1{\bar\psi}}$. 
This equation matches the geodesic of a free particle given in \eqref{FreeFall} if and only if $c_1$ vanishes. 
As a result we obtain the solution in the case with $y=0$ as 
\beal{
h(r)|_{y=0}
=&\frac{2 \arcsinh\left(\frac{H_0 }{\sqrt{\bar m}}r^{\frac{d-1}{2}}\right)}{H_0(d-1)} 
-\frac{2 w_{\parallel} r^{\frac{d-1}{2}} \Gamma \left(\frac{d+2 w_{\parallel}-3}{2 (d-1)}\right) \, _3{F}_2\left(\frac{1}{2},\frac{d+2 w_{\parallel}-3}{2 (d-1)},\frac{1}{2};\frac{-3 d-2 w_{\parallel}+5}{2-2 d},\frac{3}{2};-\frac{H_0^2 r^{d-1}}{\bar m}\right)}{(d-1)^2 \sqrt{\bar m}\Gamma \left(\frac{d+2 w_{\parallel}-5}{2 (d-1)}\right)}. 
\label{soly0}
} 
Notice that the function $h(r)$ is monotonically increasing, which is also confirmed from $h' = -\frac1\psi= ({m(t,r)\over r^{d-3}} +H_0^2 r^2 )^{-\half}$.
Note that the gravitational evaporation can happen but the gravitational collapse does not happen when the space expands exponentially in this model as in the case with a perfect fluid inside studied in the previous section. 

Let us comment on a solution of \eqref{transverse} when the cosmological constant vanishes and the dimensions are four. In this situation \eqref{transverse} can be solved in terms of $h'(r)$ by using some hypergeometric functions. The integration constant can be determined by requesting the solution to reduce in the limit $y\to0$ to the expression of \eqref{soly0} with $H_0\to0$. The result is 
\beal{
h'(r)|_{H_0=0} = 
\frac{\sqrt{r} \, _1{F}_1\left(\frac{4}{3};\frac{2 (w_{\parallel}+2)}{3};-\frac{2 r^{3/2} y}{3 \sqrt{\bar m}}\right)}{ \sqrt{\bar m} (2 w_{\parallel}+1) \, _1{F}_1\left(\frac{1}{3};\frac{2 w_{\parallel}+1}{3};-\frac{2 r^{3/2} y}{3 \sqrt{\bar m}}\right)}. 
}
Note that we could not integrate this further.

\subsection{Longitudinal component} 
\label{longitudinal} 

We first consider the case that the pressure in the equation of state consists only of the longitudinal one, $p_{\perp}=w_\perp\rho $, which is imposed at a layer specified by $t=h(r)+\Delta(1-\sigma-\varepsilon)$. 
Employing \eqref{rhopp} we can rewrite this as $
{\dot m(t,r)/\psi} = (w_\perp+1) m(t, r)'.$
We can compute this further by employing \eqref{m} as 
$
h' = -1/(w_\perp+1){\bar\psi},
$
where ${\bar\psi}$ is given below \eqref{transverse}.  
This can be solved as
\be
h(r)= c_1+\frac{ 2 \arcsinh\left(\frac{H_0 }{\sqrt{\bar m}}r^{\frac{d-1}{2}}\right)}{(d-1) H_0 (w_\perp+1)}.
\ee
We fix an integration constant $c_1$ by requesting this solution to reduce to the one for the transverse pressure \eqref{soly0} in the pressureless limit $w_\perp,w_\parallel\to0$. The result is $c_1=0$.

\subsection{Mixed pressure} 
\label{Mixed} 

Finally we consider the general case \eqref{EOSgeneral}. 
Using \eqref{rhopp} and \eqref{m} we can rewrite this as follows. 
\beal{
-yh' (h'  + \frac{1}{\bar\psi}) + (h''  - \frac{\bar\psi'}{\bar\psi^2}) +\frac{v}{r\bar\psi} =-\frac {w}r h' ,
\label{mixed}
}
where $v = (d-2)v_0$, $w = (d-2)(w_0+v_0)$, $y$ is defined below \eqref{transverse}. 
As in subsection \ref{Transverse} we solve this in the case with $y=0$. 
In this situation the above differential equation reduces to 
\beal{
(h''  - \frac{\bar\psi'}{\bar\psi^2}) +\frac{v}{r\bar\psi} =-\frac {w}r h',
}
We first solve this in terms of $h'(r)$: 
\beal{
h'(r) |_{y=0}
=& c_1 r^{-w}-\frac{r^{\frac{d-3}{2}} \left(2 (w-v) \sqrt{\frac{H_0^2 r^{d-1}}{\bar m}+1} \, _2F_1\left(\frac{1}{2},\frac{d+2 w-3}{2 (d-1)};\frac{-3 d-2 w+5}{2-2 d};-\frac{H_0^2 r^{d-1}}{\bar m}\right)-d-2 w+3\right)}{(d+2 w-3) \sqrt{H_0^2 r^{d-1}+\bar m}} , 
} 
where $c_1$ is an integration constant. Requesting this solution to reduce to the previous one in the limit $v_0\to0$ we can fix it as $c_1=0$. 
We can further integrate it as follows. 
\beal{ 
h(r)|_{y=0}
=&\frac{2 \arcsinh\left(\frac{H_0 }{\sqrt{\bar m}}r^{\frac{d-1}{2}}\right)}{(d-1) H_0}
-\frac{2 \sqrt{\pi } r^{\frac{d-1}{2}} (w-v) \Gamma \left(\frac{-3 d-2 w+5}{2-2 d}\right) \, _3\tilde{F}_2\left(\frac{1}{2},\frac{d+2 w-3}{2 (d-1)},\frac{1}{2};\frac{-3 d-2 w+5}{2-2 d},\frac{3}{2};-\frac{H_0^2 r^{d-1}}{\bar m}\right)}{(d-1) \sqrt{\bar m} (d+2 w-3)}. 
\label{soly0mix}
}

When the cosmological constant vanishes and the dimensions are four, \eqref{mixed} can be solved in terms of $h'(r)$ by using some hypergeometric functions. We fix an integration constant by requesting the solution to reduce in the limit $y\to0$ to \eqref{soly0mix} with $H_0\to0$. The result is 
\beal{
h'(r)|_{H_0=0} = 
\frac{\sqrt{r} (2 v+1) \, _1{F}_1\left(\frac{2 (v+2)}{3};\frac{2 (w+2)}{3};-\frac{2 r^{3/2} y}{\sqrt{\bar m}}\right)}{3 \sqrt{\bar m}(2w+1) \, _1{F}_1\left(\frac{2 v+1}{3};\frac{2 w+1}{3};-\frac{2 r^{3/2} y}{3 \sqrt{\bar m}}\right)}.
}

\bibliographystyle{utphys}
\bibliography{collapse}

\end{document}